# Superconductivity in SmFe$_{1-x}$M$_x$AsO (M = Co, Rh, Ir)

Yanpeng Qi, Lei Wang, Zhaoshun Gao, Dongliang Wang, Xianping Zhang, Zhiyu Zhang, Yanwei Ma[*]

Key Laboratory of Applied Superconductivity, Institute of Electrical Engineering,

Chinese Academy of Sciences, P. O. Box 2703, Beijing 100190, China

**Abstract:**

In this paper we report the comparative study of superconductivity by 3d (Co), 4d (Rh), 5d (Ir) element doping in SmFeAsO. X-ray diffraction patterns indicate that the material has formed the ZrCuSiAs–type structure with a space group P4/nmm. It is found that the antiferromagnetic spin-density-wave (SDW) order in the parent compounds is rapidly suppressed by Co, Rh, and Ir doping, and superconductivity emerges. Both electrical resistance and magnetization measurements show superconductivity up to around 10 K in SmFe$_{1-x}$M$_x$AsO (M = Co, Rh, Ir). Co, Rh and Ir locate in the same column in the periodic table of elements but have different electronic band structure, so comparative study would add more ingredients to the underlying physics of the iron-based superconductors.

---

[*] Author to whom correspondence should be addressed; E-mail: ywma@mail.iee.ac.cn

# 1. Introduction

The discovery of superconductivity at the high temperature of 26 K in LaFeAsO$_{1-x}$F$_x$ [1] last year has led to a great breakthrough in the research of high temperature superconductivity. Soon La has been replaced by other rare-earths such as Ce, Nd, Pr, Sm to yield new superconductor with the transition temperature above 50 K [2-7], which is the highest among materials except copper-oxides. These compounds have a two-dimensional ZrCuSiAs type structure (P4/nmm) in which FeAs forms a spatial network similar to the CuO plane in cuprates case. Nevertheless, although the crystal structure is substantially different from that of cuprate superconductors, both compounds share intriguing similarities (layered structure, doping-induced superconductivity, strong electronic correlations, and proximity of magnetic phases...). However, it is never stopped searching for new superconductors. Later the oxygen-free iron arsenide compounds $Ae$Fe$_2$As$_2$ (denoted as FeAs-122, where $Ae$ = Ba, Sr, Ca, Eu) [8-13] and LiFeAs [14] were found to exhibit superconductivity.

Actually, the discovery of high-temperature superconductivity in iron-based pnictides has attracted a great deal of attention, not only because of their high critical temperatures, but also because these compounds contain iron (Fe) element, which is typical magnetic element and unfavorable for superconductivity with singlet pairing. Furthermore, band structure calculation indicates that five 3d orbitals of Fe atoms contribute to the multiple Fermi surfaces [15], so understanding the role of Fe atoms in these compounds not only helps to unveil the nature of interplay between magnetism and superconductivity, but also offers opportunity to study the origin of superconductivity from transition metal d-band electrons. Chemical substitution is an effective method for altering the density of states at the Fermi level, and recently superconductivity in Co-doped SmFeAsO was firstly reported by our group [15], Co, Rh and Ir locate in the same column in the periodic table of elements, Therefore, it seems intriguing to know whether it is possible to induce superconductivity in Sm-1111 family by substituting Fe ions with Rh or Ir, if so, are there any differences or similarities between different dopant cases? Therefore, systemic and comparative study of different d-band element doping in superconducting layer is helpful to understand the underlying physics in the iron based superconductors. In this paper, we report the observation of bulk superconductivity in SmFe$_{1-x}$M$_x$AsO (M = Co, Rh, Ir). Similar to Co-doping case, Rh or Ir-doping strongly destroys the anomaly in the SmFeAsO compounds and induces superconductivity at round 10 K.

**2. Experimental**

Polycrystalline samples of SmFe$_{1-x}$M$_x$AsO (M = Co, Rh, Ir) were synthesized by one-step solid state reaction method using Sm, As, Fe, Fe$_2$O$_3$ and M (M = Co, Rh, Ir) as starting materials. The details of fabrication process are described elsewhere [6, 16]. The raw materials were accurately weighed according to the stoichiometric ratio of SmFe$_{1-x}$M$_x$AsO (M = Co, Rh, Ir), then thoroughly grounded and encased into pure Nb tubes. After packing, this tube was subsequently rotary swaged and sealed in a Fe tube. The sealed samples were heated to 1180 $^o$C and kept at this temperature for 45 hours. The high purity argon gas was allowed to flow into the furnace during the heat-treatment process. It is noted that the sample preparation process except for annealing was performed in a glove box under high pure argon.

The x-ray diffraction measurement was performed at room temperature with Cu-K$_\alpha$ radiation from 10° to 80° with a step of 0.01°. The analysis of x-ray powder diffraction data was done and the lattice constants were derived. The ac susceptibility of the samples was measured on the Maglab-12 T (Oxford) with an ac field of 0.1 Oe and a frequency of 333 Hz. The resistivity data were obtained using a four-probe technique on the Quantum Design instrument physical property measurement system (PPMS).

**3. Results and discussion**

Figure 1 shows the x-ray diffraction patterns of the samples of SmFe$_{0.9}$Co$_{0.1}$AsO, SmFe$_{0.9}$Rh$_{0.1}$AsO, SmFe$_{0.825}$Ir$_{0.175}$AsO, which have the highest superconducting transition temperature in their own systems. It is seen that all main peaks can be well indexed based on the ZrCuSiAs tetragonal structure with the space group P4/nmm, indicting that the samples are nearly single phase. There are still some small peaks coming from the second phase, as marked by the asterisks. Further analysis indicates that this tiny amount of impurity is most probably the FeAsO. By fitting the data to the structure calculated with the software X'Pert Plus, we got the lattice constants. In table 1, we show *a*-axis and *c*-axis lattice parameters for the $SmFe_{1-x}M_xAsO$ (M = Co, Rh, Ir) samples. It is clear that M-doping (M = Co, Rh, Ir) leads to an apparent decrease in c-axis lattice while the a-axis increases a bit. Similar behavior is observed in the $SrFe_{2-x}M_xAs_2$ and $LaFe_{1-x}Ir_xAsO$ compounds [17-20]. Compared to the parent compound SmFeAsO, the apparent variation of the lattice parameters upon M-doping indicates a successful chemical substitution in $SmFe_{1-x}M_xAsO$ compounds.

Figure 2 shows the temperature dependence of the electrical resistivity for $SmFe_{1-x}M_xAsO$ (M = Rh, Ir) samples in the temperature range from 300 to 2 K (electrical resistivity of $SmFe_{1-x}Co_xAsO$ not shown here). The inset shows an enlarged plot of ρ versus T at the low temperature. It is known that the undoped SmFeAsO sample exhibits a clear anomaly near 150 K [2], which is ascribed to the spin-density-wave (SDW) instability and structural phase transitions from tetragonal to orthorhombic symmetry. As seen from the Fig. 2(a), by doping Rh, the SDW transition is suppressed and the resistivity drop was converted to an uprising at lower temperature, the superconductivity ∼ 6.5 K appears in the sample with nominal composition of x = 0.05. At higher Rh-doping, the uprising in the lower temperature is not obvious, and the highest transition temperature 9.5 K is observed at x = 0.10. This is similar to the case of Co doping in the iron oxyarsenides compounds [21, 22]. With further Rh-doping, the transition temperature decreases (4.8 K for x = 0.15) and disappears again at x = 0.20. Similar to $SmFe_{1-x}Rh_xAsO$, the resistivity of $SmFe_{1-x}Ir_xAsO$ samples changes from semiconductorlike to superconductive with the Ir doing (see Fig. 2(b)), and the highest transition temperature 10.8 K is appeared at x = 0.175. Our data demonstrate that similar to Co-doping, both 4d (Rh) and 5d (Ir) elements doping could suppress the SDW transition and induce superconductivity in the Sm-1111 system. However, our preliminary results explicitly indicate the feasibility of inducing superconductivity by higher d elements in Sm-1111 family, and the superconducting transition temperature would be improved further by optimization of doping.

We measured the ac magnetic susceptibility to further confirm the superconductivity of SmFe$_{1-x}$M$_x$AsO (M = Co, Rh, Ir). Figure 3 shows the temperature dependence of ac magnetization for SmFe$_{0.9}$Co$_{0.1}$AsO, SmFe$_{0.9}$Rh$_{0.1}$AsO, and SmFe$_{0.825}$Ir$_{0.175}$AsO sample. The samples show a well diamagnetic signal and superconductivity with T$_c$ =13.6 K for SmFe$_{0.9}$Co$_{0.1}$AsO, T$_c$ =6.8 K for SmFe$_{0.9}$Rh$_{0.1}$AsO, and T$_c$ =10.5 K for SmFe$_{0.825}$Ir$_{0.175}$AsO, which are corresponding to the middle transition point of resistance, a superconducting volume fraction is large enough to constitute bulk superconductivity.

Figure 4 shows the temperature dependence of resistivity for SmFe$_{0.9}$Co$_{0.1}$AsO, SmFe$_{0.9}$Rh$_{0.1}$AsO and SmFe$_{0.825}$Ir$_{0.175}$AsO samples under different magnetic fields. Similar to other iron based superconductors, applied magnetic field is observed to suppress the transition. It is clear that the onset transition temperature is not sensitive to magnetic field, but the zero resistance point shifts more quickly to lower temperatures. We tried to estimate the upper critical field ($H_{c2}$) and irreversibility field ($H_{irr}$), using the 90% and 10% points on the resistive transition curves. The change of transition temperature ($T_c$) with critical field (H) is shown in the inset of Fig. 4. The Slope of $-$ dH$_{c2}$/dT|$_{Tc}$ is 2.1 T / K for SmFe$_{0.9}$Co$_{0.1}$AsO, 2.5 T / K for SmFe$_{0.9}$Rh$_{0.1}$AsO and 2.5 T / K for SmFe$_{0.825}$Ir$_{0.175}$AsO, respectively. From this figure, using the Werthamer-Helfand-Hohenberg formula [23], $H_{c2}(0) = 0.693 \times (dH_{c2} / dT) \times T_c$, we can get $H_{c2}(0)$ is about 20.3 T for SmFe$_{0.9}$Co$_{0.1}$AsO, 18.2 T for SmFe$_{0.9}$Rh$_{0.1}$AsO, and 16.9 T for SmFe$_{0.825}$Ir$_{0.175}$AsO, respectively. If adopting a criterion of 99 %$\rho_n$(T) instead of 90%$\rho_n$(T), the $H_{c2}(0)$ value of this sample obtained by this equation is even higher.

The discovery of superconductivity in the iron based compounds has stimulated a massive experimental and theoretical effort to uncover the mechanisms responsible for this novel superconductivity; however, the superconductivity mechanism in these new superconductors remains unclear yet. It is interesting that superconductivity was realized by doping magnetic element cobalt into the superconducting-active FeAs layers, and our results indicate that not only 3d (Co) element, but also 4d (Rh), 5d (Ir) element doping could suppresse the SDW transition and induce superconductivity. Co, Rh and Ir locate in the same column in the periodic table of elements; however, they have different electronic band structure and masses. All the three elements could suppress the SDW transition and induce similar values of the maximum superconducting transition temperatures, so comparative study would give us the important clues to understanding the superconducting mechanisms on iron-based superconductors.

**4. Conclusions**

In summary, we have successfully fabricated a series of new superconductor $SmFe_{1-x}M_xAsO$ (M = Co, Rh, Ir) by replacing the Fe with the 3d (Co), 4d (Rh), 5d (Ir) elements. X-ray diffraction patterns indicate that the material has formed the ZrCuSiAs–type structure. The presence of zero resistance and diamagnetism in the measurement proves that M (M = Co, Rh, Ir) substitution in the SmFeAsO compounds lead to superconductivity. Furthermore, through measuring the electrical resistivity under different magnetic field, we found that the superconductivity in all the doped samples is rather robust against the magnetic field with a slope of $-dH_{c2}/dT = 2.5$ T / K near Tc. Considering the location of Co, Rh, Ir in the same column in the periodic table of elements, comparative study is helpful to understand the underlying physics in the iron based superconductors.

**Acknowledgments**

The authors thank Profs. Haihu Wen, Liye Xiao and Liangzhen Lin for their help and useful discussions. This work is partially supported by the National '973' Program (Grant No. 2006CB601004) and Natural Science Foundation of China (Grant No: 50777062 and 50802093).

**Captions**

Figure 1 XRD patterns of the SmFe$_{1-x}$M$_x$AsO (M = Co, Rh, Ir) samples. The impurity phases are marked by *.

Figure 2 Temperature dependence of resistivity for the samples: (a) SmFe$_{1-x}$Rh$_x$AsO, (b) SmFe$_{1-x}$Rh$_x$AsO. Inset: Enlarged view of low temperature, showing superconducting transition.

Figure 3 Temperature dependence of DC magnetization for the SmFe$_{1-x}$M$_x$AsO (M = Co, Rh, Ir) sample.

Figure 4 Temperature dependence of resistivity for sample (a) SmFe$_{0.9}$Co$_{0.1}$AsO, (b) SmFe$_{0.9}$Rh$_{0.1}$AsO, (c) SmFe$_{0.825}$Ir$_{0.175}$AsO at different magnetic fields. Inset: The upper critical field H$_{c2}$ and H$_{irr}$ as a function of temperature for SmFe$_{1-x}$M$_x$AsO (M = Co, Rh, Ir) samples.

Table I  $a$- and $c$-axis lattice constant of SmFe$_{0.9}$Co$_{0.1}$AsO, SmFe$_{0.9}$Rh$_{0.1}$AsO, and SmFe$_{0.825}$Ir$_{0.175}$AsO samples.

| Nominal | $a$ (Å) | $c$ (Å) |
|---|---|---|
| SmFeAsO | 3.9391 | 8.4970 |
| SmFe$_{0.9}$Co$_{0.1}$AsO | 3.9412 | 8.4802 |
| SmFe$_{0.9}$Rh$_{0.1}$AsO | 3.9427 | 8.4501 |
| SmFe$_{0.825}$Ir$_{0.175}$AsO | 3.9521 | 8.4190 |

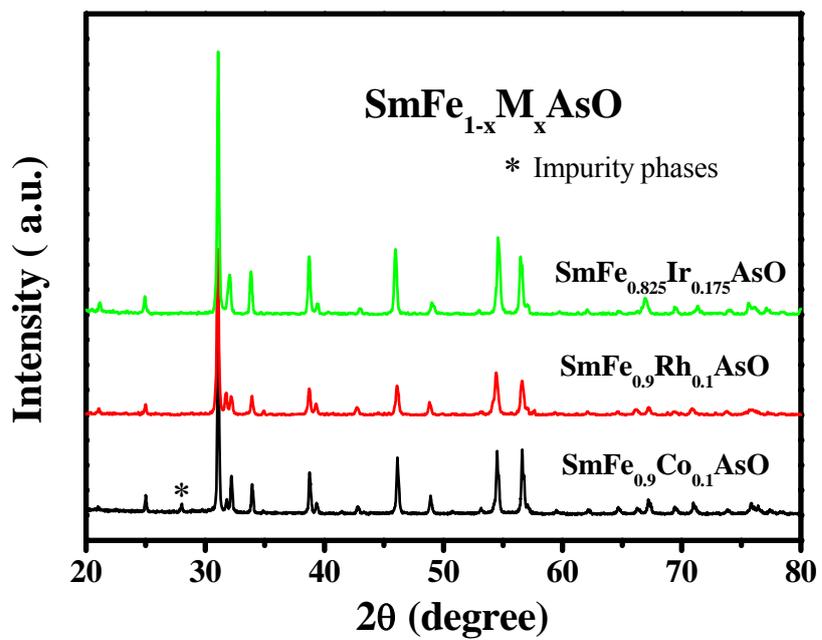

Fig.1 Qi et al.

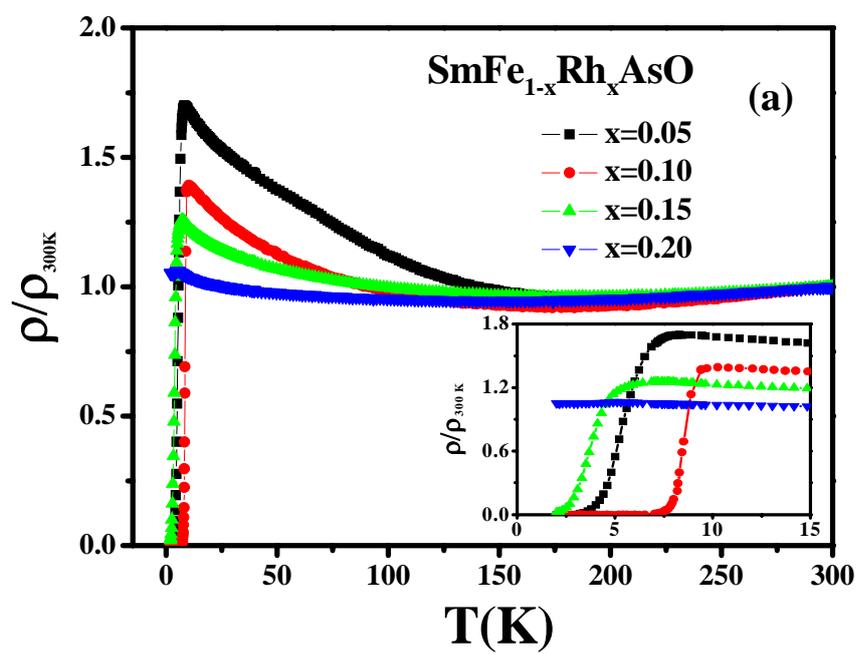

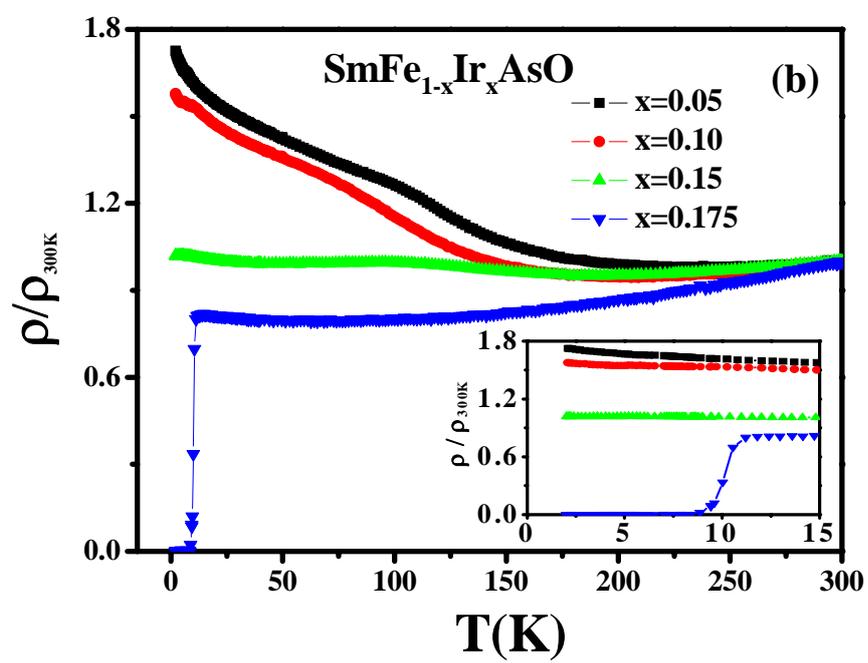

Fig.2 Qi et al.

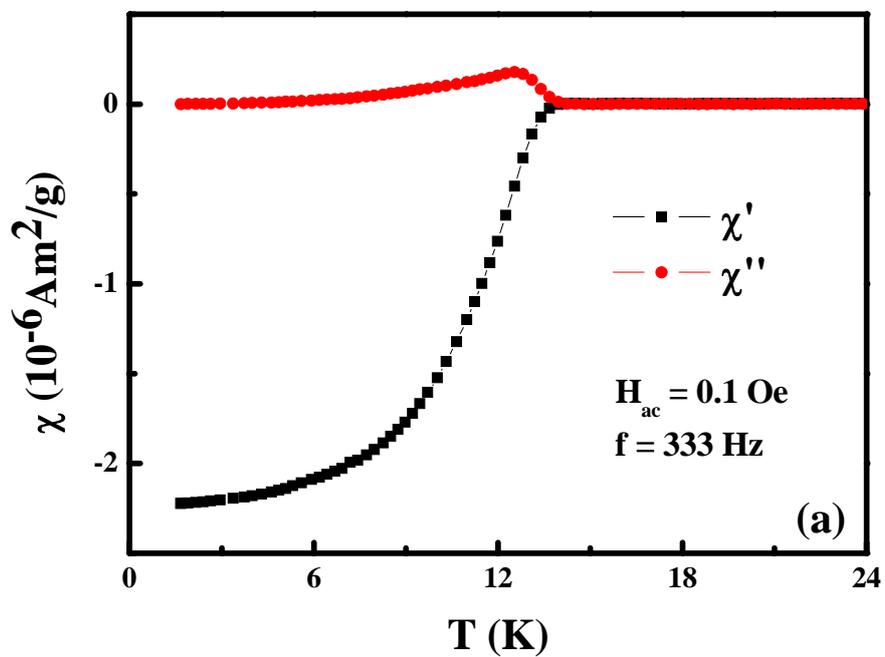

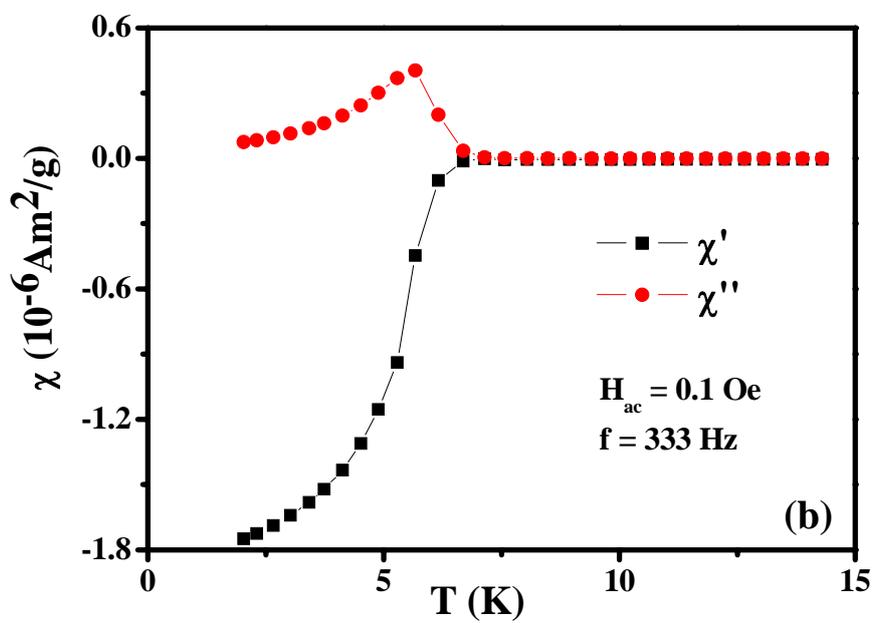

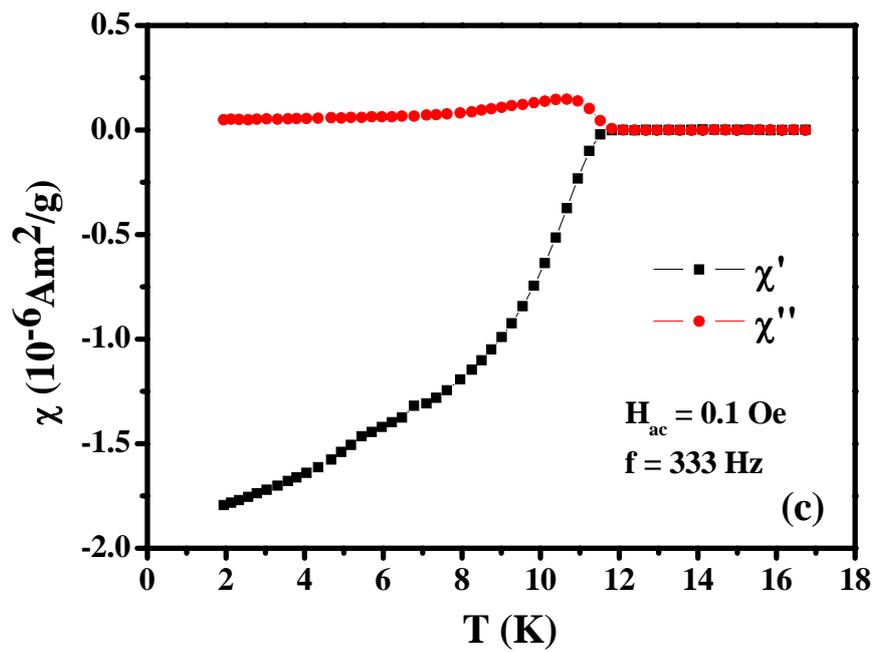

Fig.3 Qi et al.

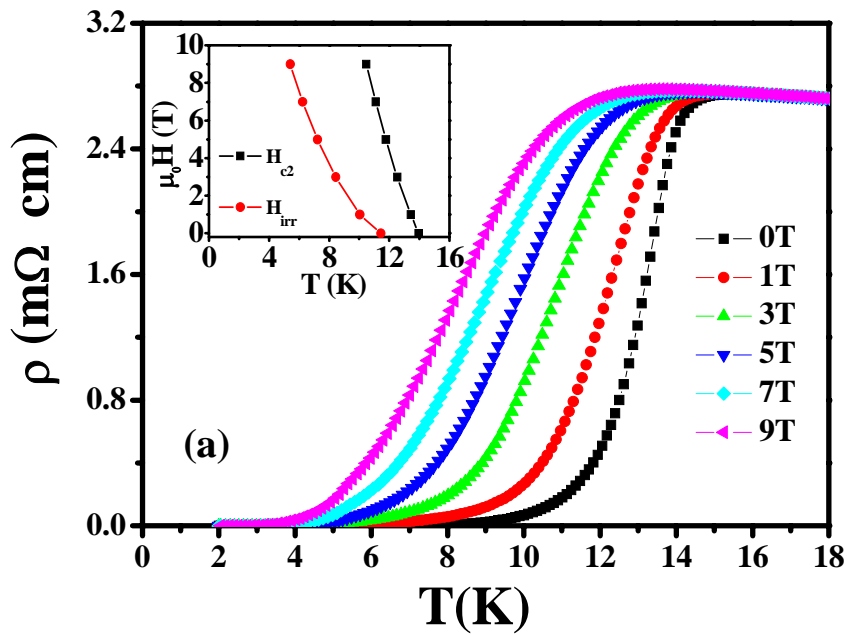

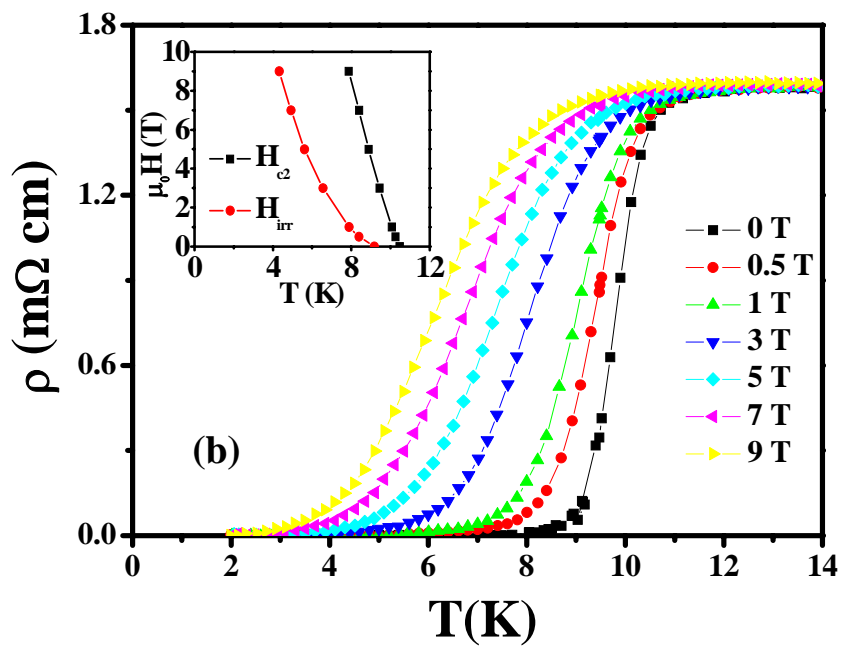

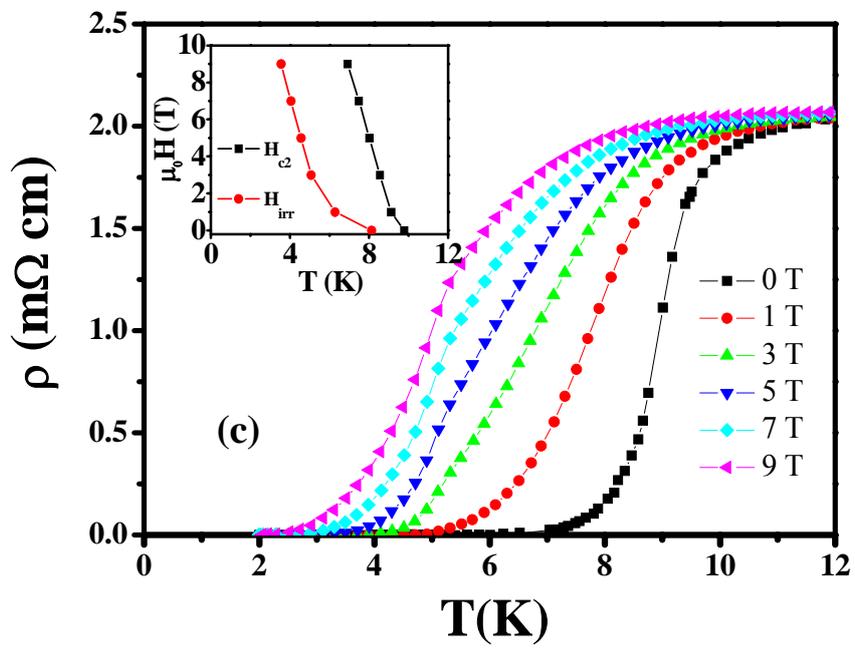

Fig.4 Qi et al.